\documentclass{PoS}

\usepackage{graphicx}

\title{Magnetic fields during primordial star formation}

\ShortTitle{Magnetic fields during primordial star formation}

\author{\speaker{Dominik R. G. Schleicher}\\
        Leiden Observatory, Leiden University, P.O. Box 9513, NL-2300 RA Leiden, the Netherlands;\\
        ESO, Karl-Schwarzschild-Strasse 2, 85748 Garching bei M\"unchen, Germany\\
        E-mail: \email{dschleic@eso.org}
        }

\author{Sharanya Sur, Robi Banerjee, Ralf S. Klessen\\
        Zentrum f\"ur Astronomie der Universit\"at Heidelberg, Institut f\"ur Theoretische Astrophysik, Albert-Ueberle-Str.~2, 69120 Heidelberg, Germany
        }

\author{Christoph Federrath   \\  
        Ecole Normale Sup\'{e}rieure de Lyon, CRAL, 69364 Lyon, France
        }
        
        \author{Tigran Arshakian, Rainer Beck\\
        Max-Planck-Institut f\"ur Radioastronomie, Auf dem H\"ugel 69, 53121 Bonn, Germany
        }

\author{Marco Spaans\\
       Kapteyn Astronomical Institute, University of Groningen, P.O. Box 800, 9700 AV, Groningen, the Netherlands
        }

\abstract{Recent FERMI observations provide a lower limit of $10^{-15}$~G for the magnetic field strength in the intergalactic medium (IGM). This is consistent with theoretical expectations based on the Biermann battery effect, which predicts such IGM fields already at redshifts  $z\sim10$. During gravitational collapse, such magnetic fields can be amplified by compression and by turbulence, giving rise to the small-scale dynamo. On scales below the Jeans length, the eddy turnover timescale is much shorter than the free-fall timescale, so that saturation can be reached during collapse. This scenario has been tested and confirmed with magneto-hydrodynamical simulations following the collapse of a turbulent, weakly magnetized cloud. Based on a spectral analysis, we confirm that turbulence is injected on the Jeans scale. For the power spectrum of the magnetic field, we obtain the Kazantsev slope which is characteristic for the small-scale dynamo. A calculation of the critical length scales for ambipolar diffusion and Ohmic dissipation shows that these scales are always small enough to allow significant amplification of the magnetic field by small-scale eddies. We discuss potential implications for the protostellar accretion disk, with particular focus on the magneto-rotational instability, which may change the morphology of the disk and reduce the accretion rate by a factor of a few.
}

\FullConference{Cosmic Radiation Fields: Sources in the early Universe - CRF2010,\\
		November 9-12, 2010\\
		Desy Germany}

\begin{document}

\section{The initial field strength: Upper and lower limits}
An observational lower limit on the magnetic field strength in the intergalactic medium (IGM) has been derived based on recent FERMI observations of TeV blazars \cite{Neronov10, Tavecchio10, Dolag10}. For such blazars, the TeV flux is known, and the expected GeV flux can be calculated by modeling the cascade of the high-energy particles. The expected flux is however orders of magnitudes higher than the current upper limit obtained with FERMI, unless magnetic fields deflected charged particles from the line of sight. Based on these data, a lower limit of the magnetic field strength can be obtained. In general, this lower limit depends on the assumed coherence length $L_B$ \cite{Neronov10}. For $L_B>0.1$~Mpc, the lower limit is of the order  $10^{-15}$~G, while for smaller coherence lengths, the lower limits increases as $L_B^{-0.5}$. As shown in \cite{Dolag10}, this IGM field fills more than $60\%$ of the volume.

This observational constraint is consistent with theoretical expectations based on the Biermann battery term \cite{Biermann50}. In a cosmological MHD simulation, the generation of magnetic fields was followed based on the Biermann battery effect \cite{Xu08}, finding IGM magnetic fields of $10^{-15}$~G at $z\sim10$, which may naturally explain the observed lower limits. Additional seed fields may be created by the Weibel instability in shocks \cite{Lazar09}. Even stronger magnetic fields may have been created in the Universe before recombination \cite{Grasso01, Banerjee04b}. For such cases, an upper limit of $\sim3$~nG (co-moving) has been derived from the observed reionization optical depth \cite{Schleicher08b}.

\section{Amplification during gravitational collapse}
\begin{figure}
\includegraphics[scale=0.42]{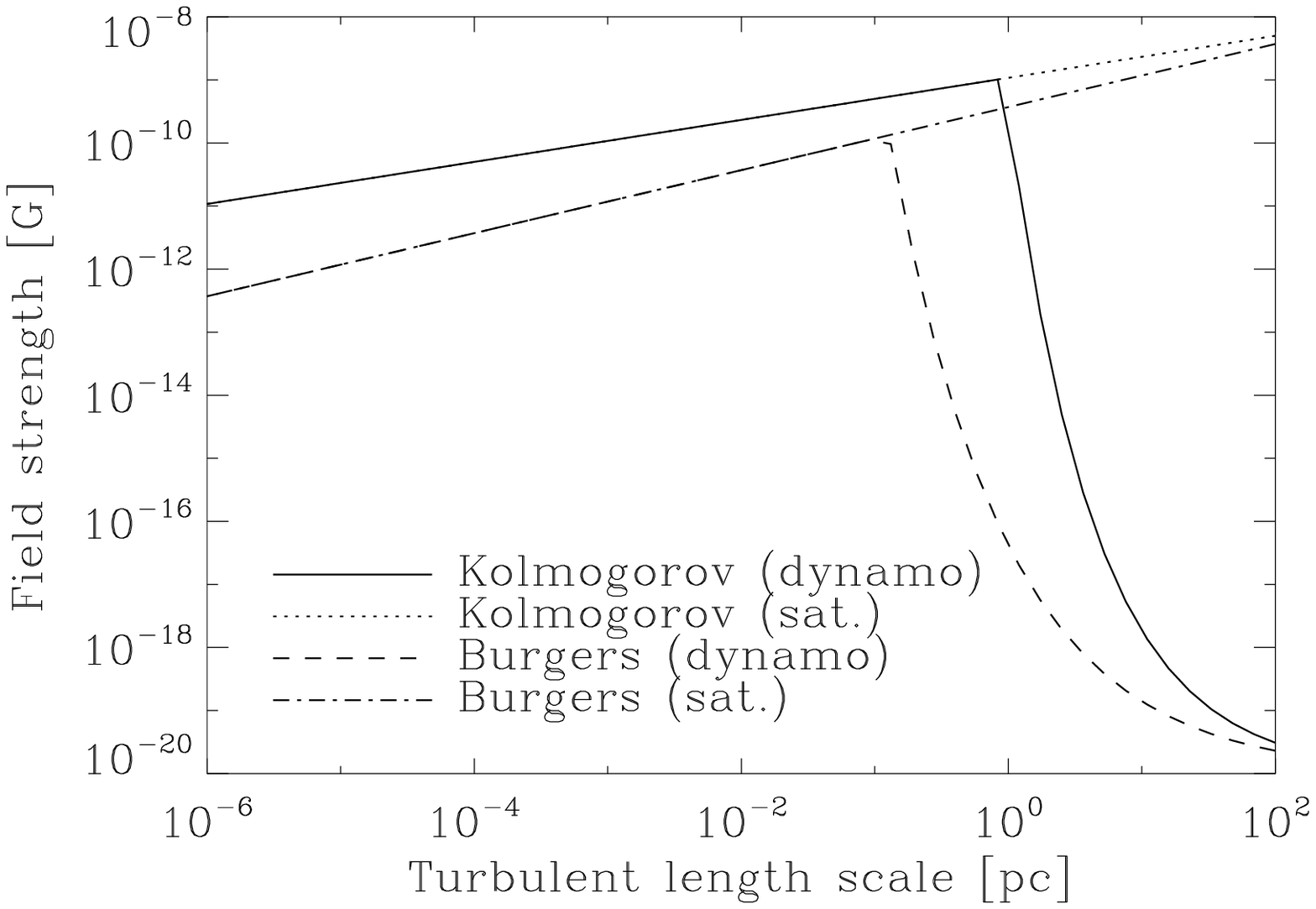}
\includegraphics[scale=0.42]{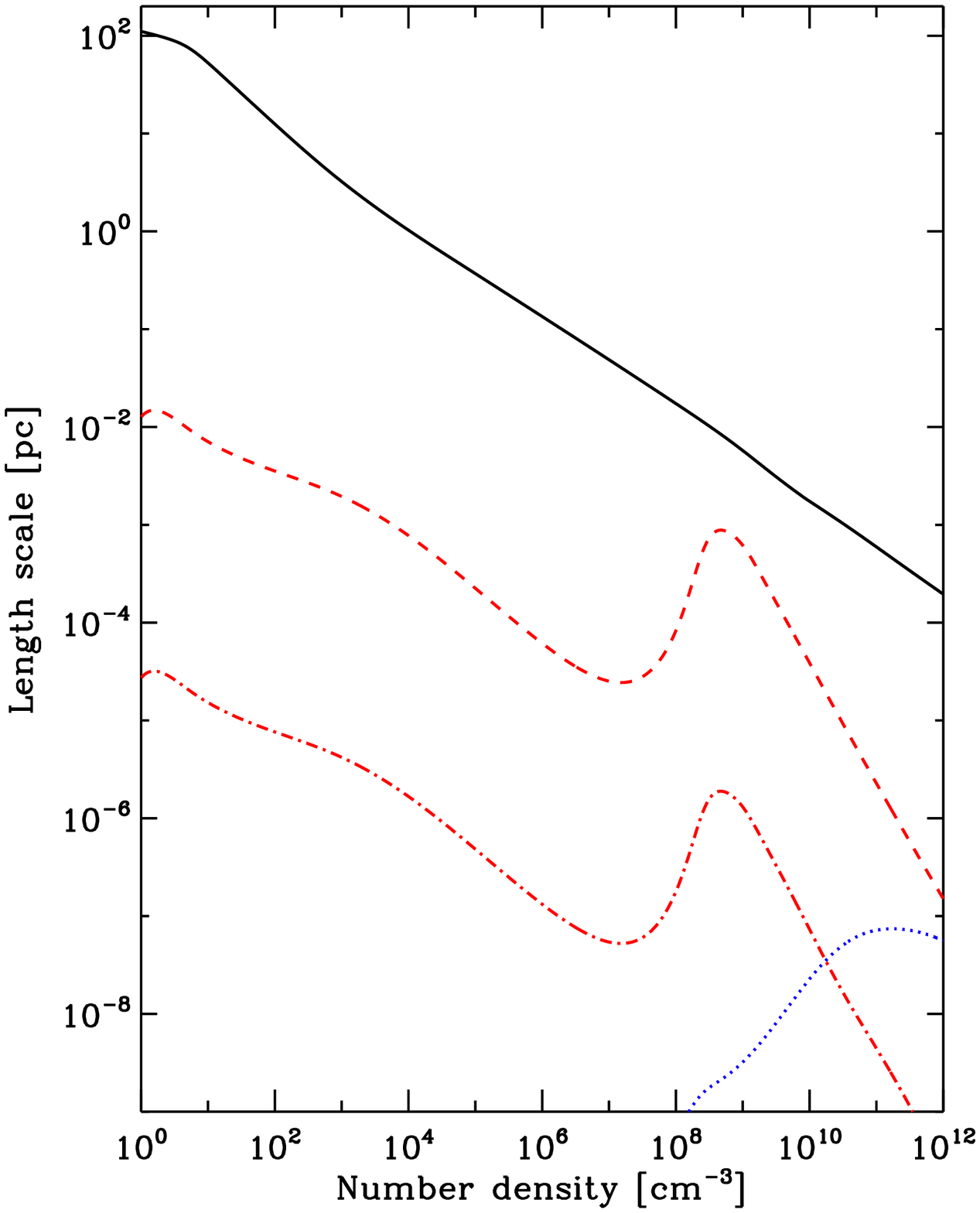}
\caption{Left: Estimates on the field strength that can be reached within a free-fall time as a function of scale. Shown are both the actual field strength to which the magnetic field is amplified by the small-scale dynamo, as well as the maximum field strength where saturation would occur. This saturation field strength increases continuously as a power-law until the scale 
where turbulence is injected, whereas the actual field strength has a maximum between $1$ and $10$~pc. \newline Right: Important length scales during gravitational collapse. The black solid line shows the Jeans length as a function of central core density. The ambipolar diffusion length scale for an equipartition magnetic field is given by the dashed red line, while the dashed-dotted line corresponds to a field at $1/60$ of equipartition. The blue dotted line gives the Ohmic diffusion scale, which is independent of magnetic field strength. In all cases, the dissipation scales are smaller than the Jeans length.}
\label{fig:spectrum}\end{figure}

During collapse and accretion, gravitational energy is released in the form of turbulence \cite{Klessen10, Elmegreen10}. Such accretion-driven turbulence can amplify weak magnetic seed fields by a process called the small-scale dynamo \cite{Schleicher10c}. For a naive estimate on the potential field amplification, we calculate the eddy-turnover timescale for Kolmogorov and Burgers type turbulence, adopting scaling laws of $v\propto l^{1/3}$ and $v\propto l^{1/2}$, respectively. On scales below $1$~pc, the eddy-timescales become considerably smaller than the collapse-timescale, so that saturation may occur during gravitational collapse. We provide estimates for the saturation field strength as well as the actual field strength that could be reached within a free-fall time in Fig.~\ref{fig:spectrum}.

This scenario has been tested using MHD simulations following the collapse of a weakly magnetized, turbulent cloud \cite{Sur10}. The simulations confirm that the magnetic field amplification is stronger than the effect of pure compression, which would lead to a scaling relation of $B\propto \rho^{2/3}$ in case of spherical symmetry. A spectral analysis shows that turbulence is injected at the Jeans scale \cite{Federrath10dyn}, and confirms the expected Kazantsev slope \cite{Brandenburg05}. Using models for ambipolar diffusion and Ohmic dissipation in primordial gas \cite{Schleicher09prim}, we checked that the resistive scales are always smaller than the Jeans scale during gravitational collapse (see Fig.~\ref{fig:spectrum}).

\section{Potential implications}
Even the presence of weak magnetic fields can change the dynamics in the protostellar accretion disk via the magneto-rotational instability \cite{Tan04, Silk06}. The interaction of the MRI with self-gravitational instabilities may effectively decrease the accretion rate and change the disk morphology \cite{Fromang04}. On the other hand, numerical simulations exploring the effects of homogeneous magnetic fields indicate the formation of jets and the suppression of fragmentation \cite{Machida08}. More detailed and self-consistent modeling is thus required to address the final implications.

The small-scale dynamo may not only produce magnetic fields during primordial star formation, but also produce a strong seed field in the first galaxies \cite{Beck94, Arshakian09, Souza10}. The presence of global rotation may lead to large-scale coherent fields already at high redshift \cite{Wang09}.

\section*{Acknowledgments}
The research leading to these results has received funding from the European Community's Seventh Framework Programme (/FP7/2007-2013/) under grant agreement No 229517 and via HPC-EUROPA2 (project number: 228398) with the support of the European Commission Capacities Area - Research Infrastructures Initiative. R. Banerjee is funded by the Emmy-Noether grant (DFG) BA 3607/1-1. RSK, SS and TGA thank for funding via {the Priority Programme 1177 {\em"Witnesses of Cosmic History:  Formation and evolution of black holes, galaxies and their environment"} of the German Science Foundation}. RSK and CF thank the German Science Foundation (DFG) for support via the Emmy Noether grant KL 1358/1, the German {\em Bundesministerium f\"{u}r  Bildung und Forschung} for subsidies via the ASTRONET project STAR FORMAT (grant  05A09VHA) and the {\em Landesstiftung Baden-W{\"u}rttemberg} for subsidies via  their program International Collaboration II.



\end{document}